\documentclass[aps,twocolumn,prd,showpacs,showkeys,preprintnumbers,superscriptaddress,nobibnotes,floatfix,longbibliography,notitlepage,nofootinbib]{revtex4-2}

\usepackage{amsmath}
\usepackage{amsfonts}
\usepackage{amssymb}
\usepackage{graphicx}
\usepackage[colorlinks=true,allcolors=blue]{hyperref}
\usepackage{relsize}
\usepackage{gensymb}
\usepackage{fontawesome}
\usepackage[capitalize]{cleveref}
\usepackage[dvipsnames,table]{xcolor}

\usepackage{siunitx}

\DeclareSIUnit \s {\second}
\DeclareSIUnit \ns {\nano\second}
\DeclareSIUnit \mus {\micro\second}
\DeclareSIUnit \ms {\milli\second}
\DeclareSIUnit \MB {\mega\byte}
\DeclareSIUnit \GB {\giga\byte}
\DeclareSIUnit \TB {\tera\byte}
\DeclareSIUnit \PB {\peta\byte}
\DeclareSIUnit \Mbps {\mega\bit/\s}
\DeclareSIUnit \Gbps {\giga\bit/\s}
\DeclareSIUnit \Tbps {\tera\bit/\s}
\DeclareSIUnit \Pbps {\peta\bit/\s}
\DeclareSIUnit \kton {\kilo\tonne} 
\DeclareSIUnit \kt {\kilo\tonne}
\DeclareSIUnit \kty {\kilo\tonne-\year}
\DeclareSIUnit \Mt {\mega\tonne}
\DeclareSIUnit \eV {\electronvolt}
\DeclareSIUnit \keV {\kilo\electronvolt}
\DeclareSIUnit \MeV {\mega\electronvolt}
\DeclareSIUnit \GeV {\giga\electronvolt}
\DeclareSIUnit \TeV {\tera\electronvolt}
\DeclareSIUnit \PeV {\peta\electronvolt}
\DeclareSIUnit \EeV {\exa\electronvolt}
\DeclareSIUnit \sr {sr}
\DeclareSIUnit \m {\meter}
\DeclareSIUnit \cm {\centi\meter}
\DeclareSIUnit \nm {\nano\meter}
\DeclareSIUnit \in {\inchcommand}
\DeclareSIUnit \km {\kilo\meter}
\DeclareSIUnit \kV {\kilo\volt}
\DeclareSIUnit \kW {\kilo\watt}
\DeclareSIUnit \MW {\mega\watt}
\DeclareSIUnit \MHz {\mega\hertz}
\DeclareSIUnit \mrad {\milli\radian}
\DeclareSIUnit \year {years}
\DeclareSIUnit \POT {POT}
\DeclareSIUnit \sig {$\sigma$}
\DeclareSIUnit\parsec{pc}
\DeclareSIUnit\lightyear{ly}
\DeclareSIUnit\foot{ft}
\DeclareSIUnit\ft{ft}
\DeclareSIUnit \ppb{ppb}
\DeclareSIUnit \ppt{ppt}
\DeclareSIUnit \samples{S}
\DeclareSIUnit \pe{PE}
\DeclareSIUnit \GeVmwe{GeV/mwe}
\DeclareSIUnit \mwe{mwe}

\newcommand{\enu}{\E_\enu}

\begin{document}

\title{Probing Quantum Gravity with Elastic Interactions of Ultra-High-Energy Neutrinos}

\author{A.~Garcia~Soto}
\affiliation{Department of Physics \& Laboratory for Particle Physics and Cosmology, Harvard University, Cambridge, MA 02138, USA}
\affiliation{Instituto de Física Corpuscular (IFIC), Universitat de València (UV), 46980 Paterna, València, Spain.}

\author{D.~Garg}
\affiliation{Department of Physics and Astronomy, University of Iowa, Iowa City, Iowa 52242, USA}

\author{M.~H.~Reno}
\affiliation{Department of Physics and Astronomy, University of Iowa, Iowa City, Iowa 52242, USA}

\author{C.~A.~Arg{\"u}elles}
\affiliation{Department of Physics \& Laboratory for Particle Physics and Cosmology, Harvard University, Cambridge, MA 02138, USA}

\date{\today}

\begin{abstract}
The next generation of radio telescopes will be sensitive to low-scale quantum gravity by measuring ultra-high-energy neutrinos.
In this work, we demonstrate for the first time that neutrino-nucleon soft interactions induced by TeV-scale gravity would significantly increase the number of events detected by the IceCube-Gen2 radio array in the EeV regime.
However, we show that these experiments cannot measure the total cross section using only the angular and energy information of the neutrino flux, unless assumptions on the underlying inelasticity distribution of neutral interactions are made.
\end{abstract}

\maketitle

\section{Introduction}
Ultra-high-energy (UHE) neutrinos provide a unique test for predictions of Standard Model (SM) interactions. 
A $\SI{0.1}\EeV$ neutrino hitting a proton, probes a center-of-mass energy ($E_{cm}$) beyond what is accessible with present colliders, opening opportunities to test the SM and a wide variety of Beyond Standard Model (BSM) scenarios in new energy regimes~\cite{Ackermann:2019cxh,Arguelles:2019rbn,Ackermann:2022rqc}. 
While UHE neutrinos are expected to be produced in the interactions of UHE cosmic rays with the cosmic microwave background~\cite{Greisen:1966jv,Zatsepin:1966jv} and a variety of astrophysical contexts (see, e.g., Refs.~\cite{Ackermann:2019ows,Ackermann:2022rqc} and references therein), they have not been detected. 
IceCube has identified neutrinos of astrophysical origin in the PeV energy range~\cite{IceCube:2020wum,IceCube:2021rpz}, yet no experiment has claimed detection of events with higher energies. 
Upper limits on the UHE neutrino flux come from underground~\cite{IceCube:2016zyt} and surface~\cite{ARA:2019wcf,Anker:2019rzo} neutrino detectors as well as from air shower experiments~\cite{PierreAuger:2019ens,ANITA:2019wyx}.

In the last decade, a number of observatories that aim to observe UHE neutrinos have been proposed. 
Using different detectable signals like electromagnetic waves from neutrino-induced showers in air (e.g., GRAND~\cite{Alvarez-Muniz:2018bhp}, BEACON~\cite{Wissel:2020fav}, TRINITY~\cite{Otte:2019aaf}, TAMBO~\cite{Romero-Wolf:2020pzh}, PUEO~\cite{PUEO:2020bnn}, POEMMA~\cite{Olinto:2020oky}) or in ice (e.g., IceCube-Gen2~\cite{IceCube-Gen2:2020qha}, RNO-G~\cite{RNO-G:2020rmc}), they will supersede current experiments, and increase the capability to detect neutrinos in the EeV regime by orders of magnitude. 

The number of neutrinos that will interact in these detectors depends on both the UHE neutrino flux and the neutrino cross section. 
Up-going events, in which the Earth shadows high-energy neutrinos, and down-going or horizontal events in ice or in the atmosphere show different sensitivities to the flux that in principle allow an extraction of the cross section~\cite{Kusenko:2001gj, Palomares-Ruiz:2005npx,Hussain:2006wg,Marfatia:2015hva}. 
In fact, recently, IceCube has reported measurements of the neutrino-nucleon cross section using through-going muons~\cite{IceCube:2017roe}, cascades~\cite{Xu:2018zyc}, and starting  events~\cite{Bustamante:2017xuy,IceCube:2020rnc}.
Similar analyses have been performed to probe the sensitivity of next-generation neutrino experiments to the neutrino-nucleon cross section~\cite{Denton:2020jft,Esteban:2022uuw,Valera:2022ylt}.
However, as we will show in this work, these reported measurements rely on assumptions on the behavior of the neutrino-nucleon cross section and are not completely model-independent measurements of it.
The assumption that the differential cross section behaves as the SM predicts, implies that these measurements cannot be used to constrain generic BSM neutrino-matter interactions that include large elastic scattering contributions to the cross section.

Specifically, a plethora of phenomenological works have investigated whether in-ice/water Cherenkov detectors would be sensitive to these new physics phenomena through proxies such as angular and energy distributions, or neutrino flavor (see, e.g., Ref.~\cite{Ackermann:2022rqc} and references therein).
A scenario that has been extensively studied in the context of UHE neutrinos~\cite{Anchordoqui:2001ei,Feng:2001ib,Uehara:2001yk,Dutta:2002ca,Alvarez-Muniz:2002snq,Jain:2002kz,Kowalski:2002gb,Kisselev:2004ze,Illana:2005pu,Anchordoqui:2006fn,Arsene:2013nca,Reynoso:2013jya,Mack:2019bps,Illana:2020jpi,Khan:2022bcl,Alok:2022xiy} involves models with extra dimensions in which the gravitational interaction may become strong at TeV energies~\cite{Arkani-Hamed:1998jmv,Randall:1999ee}. 

In this work, to illustrate the challenges of setting constraints on BSM scenarios that are far from SM-like in the inelasticity ($y$) distributions for neutrino interactions, we investigate the extent to which a TeV-scale scenario of quantum gravity can be observed using one of the new technologies, the IceCube-Gen2 radio array. 
We focus on the feature of this scenario that the BSM elastic interactions with small neutrino energy losses yield signatures distinct from commonly used angular and energy distributions.
Our analysis presents several novel factors. 
It is the first time that gravity-mediated interactions are tested using radio detectors with a detailed detector simulation. 
We explore new proxies to detect such exotic interactions, heretofore ignored in cross section analyses of UHE neutrinos, in consideration of the role of the underlying inelasticity predictions in such measurements. 
Our analysis illustrates how a BSM scenario with a cross section dominated by low-$y$ contributions evades high-energy neutrino cross section analyses~\cite{IceCube:2017roe,Xu:2018zyc,Bustamante:2017xuy,IceCube:2020rnc,Denton:2020jft,Esteban:2022uuw,Valera:2022ylt}, emphasizing the importance of case by case studies for BSM models that predict distinctly non-standard neutrino energy loss through interactions with nucleons.

\section{TeV-scale Gravity}
In scenarios with large extra dimensions, neutrinos experience transplanckian interactions with matter when the $E_{cm}$ is above the new scale of gravity.
For inelastic collisions, black hole (BH) formation and subsequent evaporation to SM particles are expected (see, e.g., Ref.~\cite{Draggiotis:2008jz} and references therein).
Whenever the neutrino and the parton interact, single BH production can be estimated with a geometric cross section.
However, a precise estimation of these processes is difficult since it is very sensitive to effects like graviton emission during the collapse and non-thermal effects in the evaporation~\cite{Yoshino:2001ik, Yoshino:2002br,Cardoso:2002pa,Anchordoqui:2003jr,Anchordoqui:2003ug,Yoshino:2005hi}.
Here, we assume that the BH cross section can be neglected, and we focus on the very large elastic contribution to the cross section.

For elastic collisions in which the scattering occurs at impact parameters larger than the BH horizon (for small momentum transfers), the process can be described by the eikonal approximation~\cite{Kisselev:2004ze,Illana:2005pu,Illana:2020jpi}. 
In eikonal scatterings, the incoming neutrino loses a small fraction $y\equiv \Delta E_\nu/E_\nu$ of its energy and starts a hadronic shower of energy $E_{sh}$ where $E_{sh} \ll E_{\nu}$.
Several studies introduce the possibility of observing new multi-bang signatures in underground detectors, in which the neutrino interacts several times inside the detector, via eikonal interactions~\cite{Illana:2005pu,Illana:2020jpi}.
For completeness, we include the expressions for the differential $\nu N$ cross section in Appendix~\ref{appendix:xsec}.

To illustrate how low-inelasticity collisions of neutrinos with nuclear targets translates to IceCube-Gen2 radio signals, we will focus on the long-distance interactions of the 5-dimensional Randall-Sundrum model of TeV gravity~\cite{Randall:1999ee} outlined in Refs.~\cite{Illana:2005pu,Illana:2020jpi}. 
It has two parameters: the scale where gravity becomes strong ($M_{5}$) and the mass of the first Kaluza-Klein (KK) excitation ($m_{c}$). 
In this calculation, the scale $M_{5}$ must be larger than the gap between consecutive KK modes (i.e., $m_c<M_5/10$). 
The introduction of the scale $m_c$ ensures that this model is consistent with classical gravity on astrophysical scales~\cite{Giudice:2004mg,Illana:2020jpi}.
To avoid problems with observations from astrophysics~\cite{Hannestad:2003yd} and cosmology~\cite{Hannestad:2001nq}, $m_{c}$ must be larger than 50~MeV. 
In addition, collider experiments set constraints on $M_{5}$ larger than $2-\SI{4}\TeV$~\cite{Giudice:2004mg,Franceschini:2011wr}.

\begin{figure}[!h]
\centering
\includegraphics[width=0.45\textwidth]{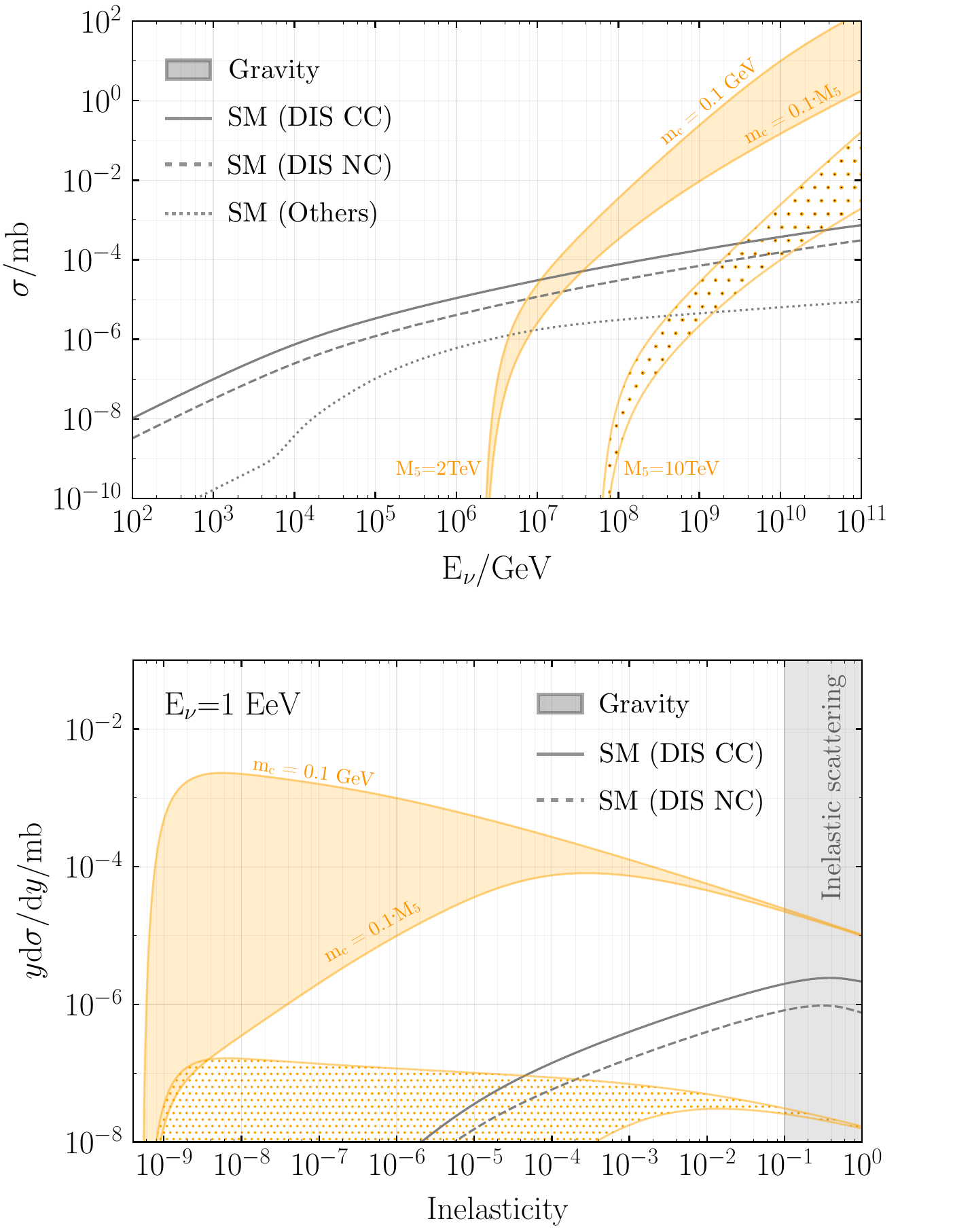}
\caption{\textbf{\textit{Standard Model and gravity-mediated cross sections.}}
Total (top) and differential (bottom) cross sections for neutrino scattering on oxygen. 
Grey lines represent the SM predictions for deep inelastic scattering (continuous and dashed) and photon-mediated interactions (dotted).
Orange bands indicate the gravity-mediated predictions assuming $M_5$ equal to $\SI{2}\TeV$ and $\SI{10}\TeV$ for different values of $m_{c}$ ranging from $\SI{0.1}\GeV$ to $M_5/10$.
The very inelastic region is shown with a grey band.}
\label{fig:xsec}
\end{figure}

Fig.~\ref{fig:xsec} shows the total and differential cross section for gravity-mediated and SM interactions. 
The TeV-gravity cross sections grow rapidly when the neutrino energy is above a threshold energy $E_{th}=M_{5}^{2}/(2m_{p})$, becoming larger than the one mediated by $W$ and $Z$ bosons.
The elasticity of these interactions strongly depends on the value of $m_{c}$. 

In summary, our focus here is on a TeV-scale gravity scenario in which a neutrino with $E_{\nu}>E_{th}$ will have a large interaction cross section; however, the neutrino interaction yields a small energy deposition relative to its initial energy as it propagates through matter, until it suffers an SM interaction.

\section{Radio Detection of TeV-scale Gravity Interactions} To study the impact of this model of TeV gravity on radio detectors, we have developed a full simulation chain.
First, we estimate the expected flux at the detector using \texttt{NuPropEarth}~\cite{Garcia:2020jwr}, where we account for attenuation effects both from SM and gravity-mediated interactions. 
The same package is used to generate the neutrino interactions in ice, where we account for multiple energy depositions due to gravity-mediated interactions. 
Finally, \texttt{NuRadioMC}~\cite{Glaser:2019cws} is used to simulate the Askaryan radio emission that comes from the shower charge excess from neutrino interactions in ice, the Askaryan radio signal propagation to the antennas, and the subsequent detector triggering.

In this work, we account not only for particle showers created by the initial neutrino interaction but also for muons and taus that might interact or decay producing a second spatially displaced shower that generates Askaryan radiation. 
The lepton propagation and generation of these secondary particles (mainly due to bremsstrahlung, pair production, and photonuclear interactions) is done using
\texttt{PROPOSAL}~\cite{Koehne:2013gpa,Dunsch:2018nsc}.
It is crucial to include these secondary showers since they can mimic the characteristic multi-bang signature of gravity-mediated interactions.

For this study we use one of the baseline configurations that have been proposed for the IceCube-Gen2 radio array~\cite{IceCube-Gen2:2020qha}, which covers a total surface area of $\SI{500}\km^2$.
It consists of 537 stations containing four log-periodic dipole antennas (LPDA) near the surface with slim dipoles deployed deeper in the ice. 
The Askaryan emission is computed using the ARZ prescription~\cite{Alvarez-Muniz:2020ary}. 

\begin{figure}[!h]
\centering
\includegraphics[width=0.45\textwidth]{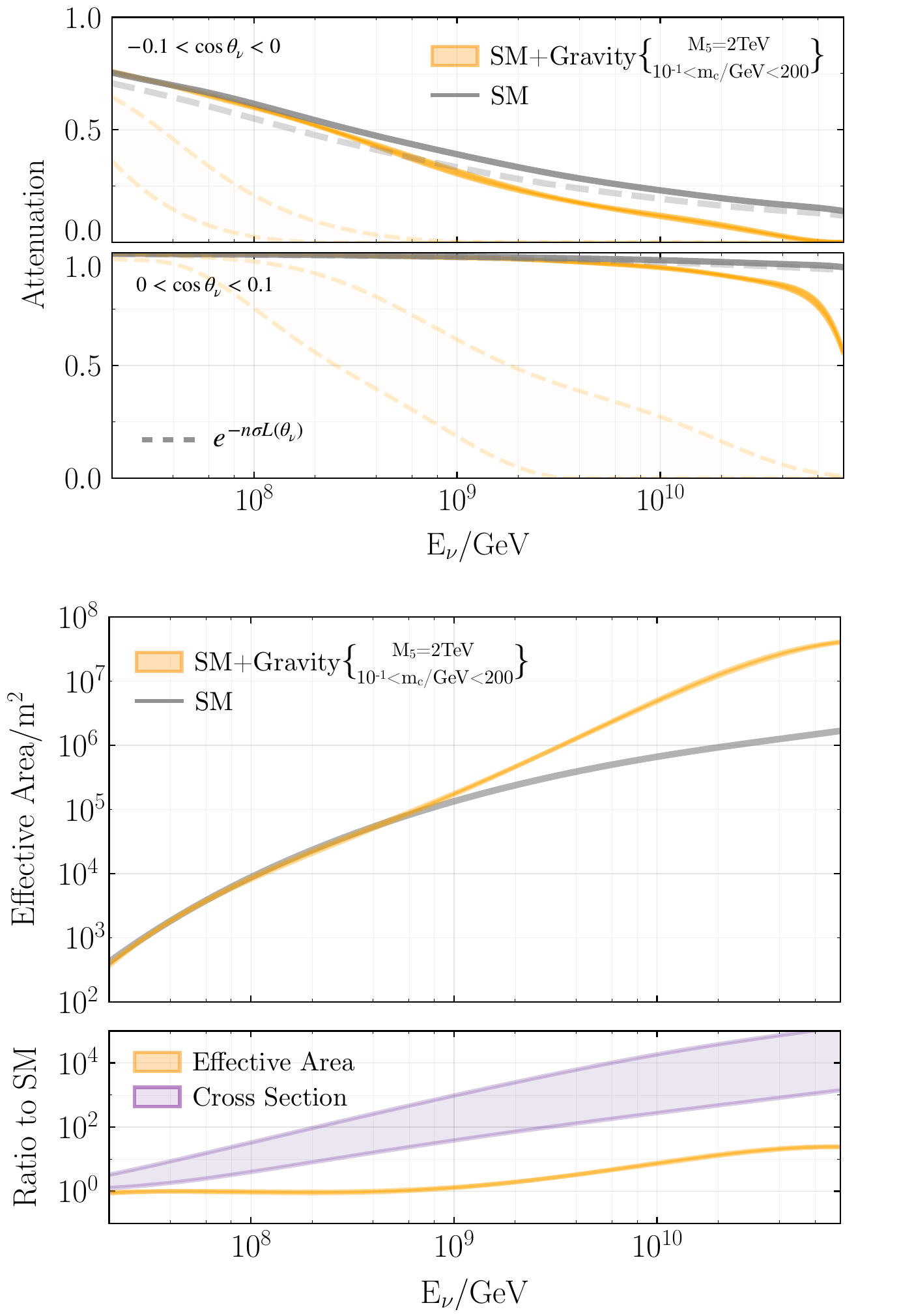}
\caption{\textbf{\textit{Attenuation and effective area.}}
Top: Attenuation of the flux as a function of the neutrino energy for events just below and above the horizon. 
Grey lines represent the expectation assuming only SM interactions. Orange band regions include SM as well as gravity-mediated interactions assuming $M_{5}=\SI{2}\TeV$ and $m_{c}=\SI{0.1}\GeV$ or $M_5/10$. Dashed lines and semitransparent bands represent the attenuation assuming neutrinos are absorbed after their first interaction.
Bottom: Effective area as a function of the neutrino energy for the IceCube-Gen2 radio array using the same color scheme of top plot, and the ratio to the SM predictions of the total cross section (purple) and effective area (orange) when gravity-mediated scattering is included.}
\label{fig:aeff}
\end{figure}

Gravity-mediated interactions play a secondary role in the flux attenuation as can be observed in Fig.~\ref{fig:aeff}, where the upper panels shows neutrino flux attenuation for angles just above and below the horizon. 
We note that this behavior is entirely different from the SM-like and commonly used behavior that depends on the number of targets along the column depth $L(\theta)$, i.e., $e^{-n\sigma L(\theta)}$, in which neutrinos are absorbed after their first interaction, shown with the light shaded region in the upper panel.
The main reason is that even though the UHE neutrinos experience many gravity-mediated interactions as they propagate through the Earth, they lose a relatively small fraction of energy per interaction. 

The effective area of the IceCube-Gen2 radio array for the SM and the SM plus gravity-mediated interactions are shown in the lower panels of Fig.~\ref{fig:aeff}.
The effective area shows a rapid growth for energies above $\SI{1}\EeV$ when $M_{5}=\SI{2}\TeV$ is assumed. 
At these energies, gravity-mediated interaction can produce hadronic showers above the detector energy threshold, even with low inelasticities. 
In the case of radio experiments, the shower must emit enough radiation to induce a pulse in the antennas above the thermal noise. For the detector configuration used in this work, the energy threshold is $\sim \SI{1}\PeV$.
Since only a fraction of the BSM cross section contributes to observable signals, the ratio of the effective areas for SM+Gravity to SM is much smaller than the ratio of SM+Gravity to SM neutrino cross sections, as shown in the lowest panel in Fig. \ref{fig:aeff}.

Furthermore, it is important to note that even though varying $m_{c}$ yields very different predictions of the total cross section, the resulting flux attenuation and effective areas are similar. 
For example, for $E_\nu=\SI{1}\EeV$ and $M_5=\SI{2}\TeV$, the BSM cross section band is a factor of $10-1300$ times larger than the SM cross section; however, the attenuation factor decreases only by $10-25\%$ and the effective area increases by a factor of $20\%$. 
This feature is not observed with SM interactions, and it is intrinsically related to the low inelasticity of the gravity-mediated interactions and the experiment-dependent detection threshold. 

\section{Proxies} We will now investigate the consequences of this effect on different observables that can be extracted with radio detectors.

{\textit{Number of Events ---}}
To begin with, we evaluate how gravity-mediated interactions would alter the expected number of events after ten years of data taking with the IceCube-Gen2 radio array. 
To do so, we convolve the effective area with three different fluxes: a single power law with spectral index given by the best-fit point found in the IceCube analysis using high-energy starting events (HESE)~\cite{IceCube:2020wum}, and two very different predictions of the GZK flux given in Ref.~\cite{vanVliet:2019cpl}.
In the three cases, we assume a democratic flavor composition of the flux at Earth.
The number of events are shown in Table~\ref{tab:rate} assuming $M_{5}=\SI{2}\TeV$.
An enhancement of the number of triggered events can be observed for both GZK flux predictions. 
In fact, detecting more than O(10) events per year cannot be explained even with optimistic scenarios of the GZK flux and would hint to gravity-mediated interactions. 
However, the uncertainty in the cosmogenic fluxes makes it infeasible to constrain any BSM scenario just by looking at the overall number of triggered events.

\begin{table}[h]
\caption{\textbf{\textit{Number of triggered events.}}
Total number of triggered events after ten years of data taking with the IceCube-Gen2 radio array for different scenarios.
Each column assumes a different neutrino flux: single power law using the best-fit point from HESE analysis (first) \cite{IceCube:2020wum}, and GZK fluxes with $10\%$ (second) and $20\%$ (third) proton component \cite{vanVliet:2019cpl}. 
First row indicate the SM predictions, whereas second and third row include the gravity-mediated contributions assuming $M_{5}=\SI{2}\TeV$ and two different values of $m_{c}$.}
\begin{tabular}{l r r r}
\toprule
 & HESE & GZK$_{20\%}$ & GZK$_{10\%}$ \\
\colrule
SM                                         & 2.0 &  671.4 & 24.6 \\
+Gr ($m_{c}=\SI{0.1}\GeV$) & 2.0 & 3971.7 & 37.2 \\
+Gr ($m_{c}=\SI{200}\GeV$)   & 2.0 & 3750.2 & 37.1 \\
\botrule
\label{tab:rate}
\end{tabular}
\end{table}

\begin{figure*}
\centering
\includegraphics[width=1\textwidth]{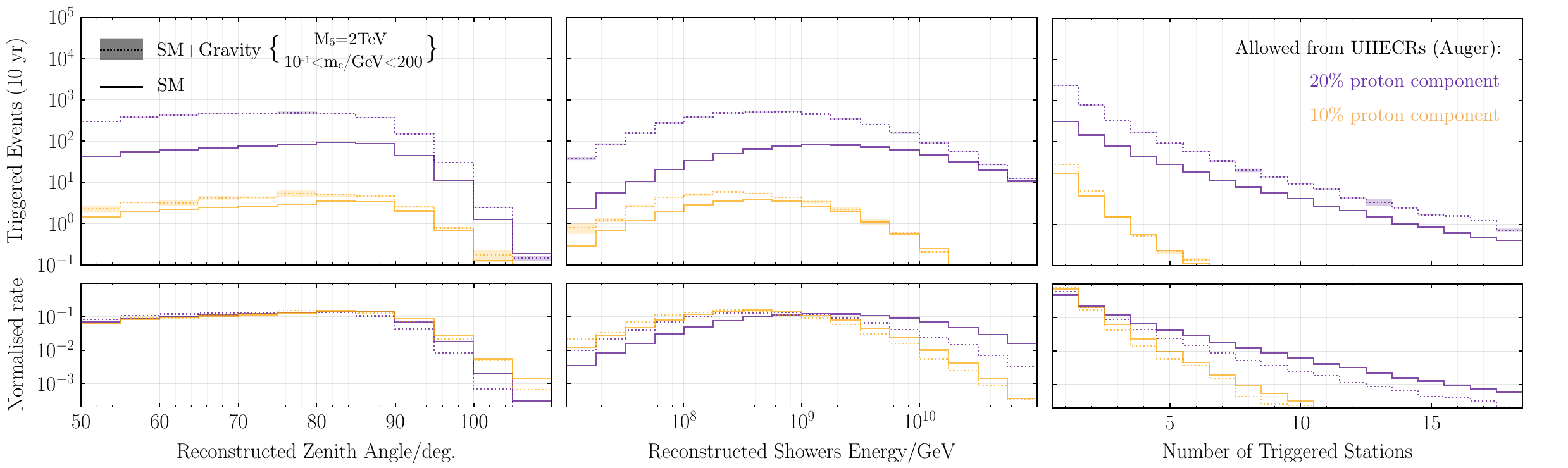}
\caption{\textbf{\textit{Observable distributions.}}
Expected zenith distribution (left), energy distribution (center), and number of triggered stations (right) after ten years of data taking with the IceCube-Gen2 radio array assuming two different GZK fluxes indicated with orange and purple colors. 
Continuous lines represent the SM prediction and shaded bands include Gravity-mediated interactions assuming $M_{5}=\SI{2}\TeV$ and $m_{c}$ ranging from $\SI{0.1}\GeV$ to $M_5/10$.}
\label{fig:rate}
\end{figure*}

\textit{Angular Distributions ---}
Previous works have shown that the angular distribution is a promising proxy to measure neutrino cross section and constrain some BSM scenarios using radio detectors~\cite{Bustamante:2017xuy,Denton:2020jft,Esteban:2022uuw,Huang:2021mki}.
Therefore, it is interesting to evaluate how gravity-mediated interactions would alter the expected SM angular distribution with the IceCube-Gen2 radio array. 
The left panel in Fig.~\ref{fig:rate} presents the expected number of triggered events as a function of the reconstructed zenith angle, which has been obtained by smearing the incoming neutrino angle with a Gaussian function with a $68\%$ width of $3^{\circ}$~\cite{Glaser:2019rxw,Valera:2022ylt}.
An enhancement of triggered events can be observed for down-going and horizontal events, whereas fewer up-going events are expected, as shown in the upper left panel of Fig.~\ref{fig:rate}. 
Since uncertainties in the GZK flux are large, we quantify the impact of this new phenomena by looking at shape differences of the angular distribution for the different scenarios, shown with the normalized rate in the lower panel.
It can be concluded that it is challenging to disentangle various flux predictions and gravity-mediated interactions using only angular information. 
In addition, as previously noted, different values of $m_{c}$ result in very different total cross sections but similar angular distributions.

\textit{Energy Depositions ---}
The main signatures of gravity-mediated elastic interactions are multiple energy depositions along the neutrino path.
Therefore, it is interesting to evaluate the impact of TeV-scale gravity elastic interactions on observables related to the energy deposition and shower multiplicity of triggered events. 
Different works have developed reconstruction techniques to infer the energy of neutrino-induced showers ($E_{dep}$)~\cite{Glaser:2022lky,Anker:2019zcx}.
Optimizations were performed to reconstruct the energy of the shower produced in the primary neutrino interaction, reporting resolutions for $\sigma\bigl(\log_{10}(E_{dep})\bigr)$ between 0.1 and 0.4. 
At the moment, there are no dedicated energy reconstructions for events with multiple showers, produced in $\nu_{\mu,\tau}$ charged current as well as gravity-mediated interactions.
Here, we assume that  $\sigma\bigl(\log_{10}(E_{dep})\bigr)=0.4$, where $E_{dep}$ is the sum of all the triggered showers produced in each event. 
The center panel of Fig.~\ref{fig:rate} shows the distributions for the sum of shower energies in our IceCube-Gen2 radio array configuration for neutrinos with SM and SM+Gravity interactions with our default $M_5=\SI{2}\TeV$ and range of $m_c$ values, for the benchmark GZK fluxes. 
While the BSM gravity interactions distort the shape of the triggered events as a function of total reconstructed shower energy, when normalized, the two GZK fluxes make differentiation between SM+Gravity and SM neutrino interactions difficult.

The right panel of Fig.~\ref{fig:rate} shows the number of triggered stations, which is correlated with the shower multiplicity of each event. 
The gravity-mediated contributions that increase the number of reconstructed events with low energy also have a large effect on the number of low-multiplicity events. 
The distributions show significant shape differences between the two flux scenarios, which create a degeneracy with TeV gravity contributions.
For these proxies, we also notice a very small impact of $m_{c}$, even though cross section predictions are extremely different. For reference, the number of triggered events as a function of neutrino zenith angle, neutrino energy and shower multiplicity are shown in the Appendix~\ref{appendix:neutrino}, as are multiplicities per neutrino flavor in the Appendix~\ref{appendix:norm_multiplicity}. We observe that events triggered by muon and tau neutrinos in SM interactions are detected by more stations than electron neutrinos. This is due to catastrophic energy losses that muons and taus can have as they propagate, which are not possible for electrons. The interesting effect of neutrino eikonal scattering is that it would distort the distribution of the number of triggered stations because it differs from electromagnetic energy losses.

\section{Conclusion and outlook} 
Radio detectors will be sensitive to neutrino energies never probed before. 
They will provide a unique opportunity to test quantum gravity scenarios in parameter space unprobed by collider experiments.
In the elastic regime, TeV-scale gravity predicts significantly larger neutrino cross sections than the SM predictions at UHE. 
Previous works showed interesting prospects to detect TeV-scale gravity, yet they lacked detailed simulation studies. 
In this manuscript, we have evaluated whether these new detector technologies will be sensitive to this BSM scenario including a detailed simulation of the neutrino interactions in the Earth and radio signal propagation and detection.

We have shown that gravity-mediated elastic interactions will significantly change the number of events triggered with the IceCube-Gen2 radio array. 
Additionally, we show that the shape of the angular distribution is only mildly affected by the new physics, whereas other distributions such as the number of triggered stations and the deposited energy are distorted by this exotic scenario. 
However, our current understanding of the UHE neutrino flux makes it challenging to disentangle the TeV gravity contributions from these distortions.
This degeneracy highlights the need for dedicated multi-cascade reconstructions in radio detectors, since so far reconstructions have only focused on single cascades. 
Furthermore, Earth-skimming neutrino experiments, which will have different sensitivities to this BSM scenario, can provide complementary information.
Finally, the flux atmospheric muons created by cosmic-ray interactions in the atmosphere can be a relevant background for down-going events with reconstructed energies below 0.1EeV. 

We have also demonstrated that these detectors are insensitive to the mass of the first KK excitation, which plays a crucial role in the prediction of the total cross section and inelasticity distribution for gravity-mediated interactions. 
This feature has an important implication in a more general context, since it demonstrates that the commonly used method to extract high-energy neutrino cross sections breaks down in scenarios where elastic interactions are significant.
While classes of BSM models dominated by large inelasticity in neutrino scattering can be analyzed in a relatively model-independent manner, this TeV-scale gravity scenario highlights the importance of model dependent considerations when distinctly non-standard $y$-distributions are important. 
Indeed, more complex BSM models, for example, the current model with the addition of a large BH cross section at ultra-high energies, will demand a fully model-dependent analysis of the cross section.

\section*{Acknowledgments}
We thank Manuel Masip and José I. Illana for useful discussions on TeV gravity; Christian Glaser, Steffen Hallmann, and Victor Valera for engaging discussions on propagation of UHE neutrinos; Ivan Esteban and John Beacom for their comments on this manuscript.
AG acknowledges support from the European Union’s H2020-MSCA Grant Agreement No.101025085.
DG and MHR are supported in part by US DOE grant DE-SC-0010113.
CAA is supported by the Faculty of Arts and Sciences of Harvard University and the Alfred P. Sloan Foundation.

\appendix

\section{Neutrino cross section with low-scale quantum gravity} \label{appendix:xsec}
The expressions for the differential $\nu N$ cross section in terms of the eikonal amplitude ${\cal A}_{\rm eik}$ and parton distribution functions $f_i(x,\mu)$ that depend on scale $\mu$ following refs. \cite{Illana:2005pu,Illana:2020jpi}:
\begin{equation}
    \frac{{\rm d}\sigma_{\rm eik}}{{\rm d}y}
    = \int_{M_5/s}^1 {\rm d}x\,
    \frac{1}{16\pi x s}
   |{\cal A}_{\rm eik}(xs,xys)|^2\sum_{q,\bar{q},G}f_i(x,\mu)
\end{equation} 
for inelasticity $y= Q^2/(xs)$ and center of mass energy squared $s=(p_\nu+p_N)^2$. The eikonal amplitude is 
\begin{eqnarray}
    {\cal A}_{\rm eik}(s,Q^2) &=& 4\pi s b_c^2 F_1 (b_c Q)\Bigl(\coth \frac{\pi Q}{m_c}- \frac{m_c}{\pi Q}\Bigr)\\
   |F_1(u)|&\simeq &1/\sqrt{1.57u^3+u^2}\\
   b_c&=&s/(2 M_5^3)\,,
\end{eqnarray}
where $m_c$ is the mass of the lightest Kaluza-Klein excitation and $M_5$ is the scale where gravity becomes strong.
The scale $\mu$ in the parton distribution functions is equal to Q for $Q<1/b_c$ and $Q^2b_c$ otherwise.

In our main text, we have considered $y<0.1$ in evaluating the cross section and neutrino showering from TeV gravity interactions to avoid over-estimating the eikonal scattering. 
We show results for a wide range of $m_c$ to make the point of the relative insensitivity of signatures in the IceCube-Gen2 radio array. Collider limits on $m_c$ are in the few TeV range, depending on $M_5$ (see, for example, ref. \cite{Landsberg:2015llu}).

Inelastic scattering of neutrinos via KK exchange can occur. Using the eikonal formula extended to $y=1$ increases the cross section, but does not change the conclusions regarding the degeneracies in the interpretations of distributions given the uncertainties in the GZK flux, as discussed below.
Inelastic scattering of neutrinos with nucleons may also occur with the production of microscopic black holes (BH). The BH cross section of neutrinos with partons depends on the neutrino-parton (neutrino-nucleon) center-of-mass energy squared $\hat{s}\ (s)$. It  
is typically written in terms of the Schwarzschild radius 
$r_H(\hat{s})\propto (\hat{s}/M_5^2)^{1/4}M_5^{-1}$. 
The geometric cross section $\hat{\sigma}_{BH}=\pi (r_H(\hat{s}))^2$ is weighted by the sum of parton distribution functions and integrated over $x=\hat{s}/s$ from $(M_{BH}^{\rm min})^2/s \to 1$. When $M_{BH}^{\rm min}=M_5$, the BH cross section can exceed the SM neutrino nucleon cross section at high energies. However, a semi-classical approach to this cross section is more suited to $M_{BH}^{\rm min}\gg M_5,$
and furthermore, energy may be lost in collisions to gravitational waves that would change signatures of microscopic black holes in detectors \cite{Yoshino:2001ik, Yoshino:2002br,Cardoso:2002pa,Anchordoqui:2003jr,Anchordoqui:2003ug,Yoshino:2005hi}.  

\section{Effect of inelastic eikonal scattering on observables} 
\begin{figure*}
\centering
\includegraphics[width=1\textwidth]{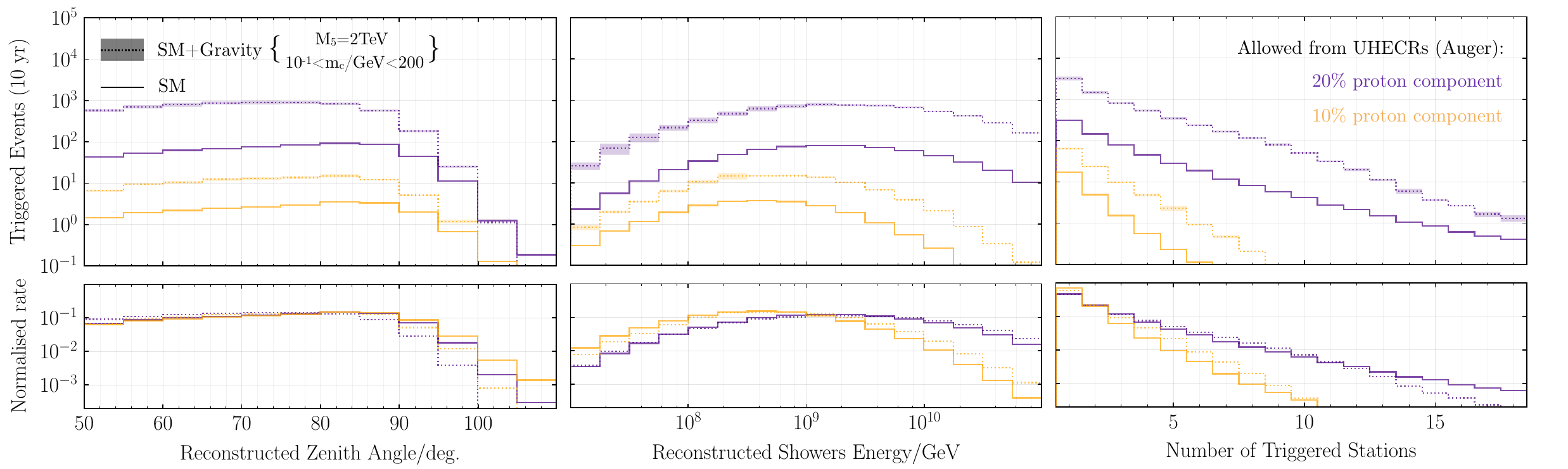}
\caption{Expected zenith distribution (left), energy distribution (center), and number of triggered stations (right) after ten years of data taking with the IceCube-Gen2 radio array assuming two different GZK fluxes indicated with orange and purple colors. 
Continuous lines represent the SM prediction and shaded bands include Gravity-mediated interactions assuming $M_{5}=\SI{2}\TeV$, $m_{c}$ ranging from $\SI{0.1}\GeV$ to $M_{5}/10$, and including the most inelastic component $0.1<y<1$.}
\label{fig:rate_ycut}
\end{figure*}

In the main text, we don't account for scattering at high inelasticities ($y>0.1$). 
Fig.~\ref{fig:rate_ycut} shows the distribution for the observables 
describe in the main text assuming that eikonal scattering works up to $y=1$ (i.e., not neglecting the inelastic contribution of this process).

The main effect of including the inelastic component is increasing the number of triggered showers at higher energies, distorting the shape of the three distributions. Noted that, the conclusions are the same as in the main text. On the one hand, eikonal scattering would have a significant effect in the Icecube-Gen2 radio array, but it is degenerated with current uncertainties in the GZK. On the other hand, there is no sensitivity to constrain $m_{c}$.

\section{Neutrino distributions} \label{appendix:neutrino}
\begin{figure*}
\centering
\includegraphics[width=1\textwidth]{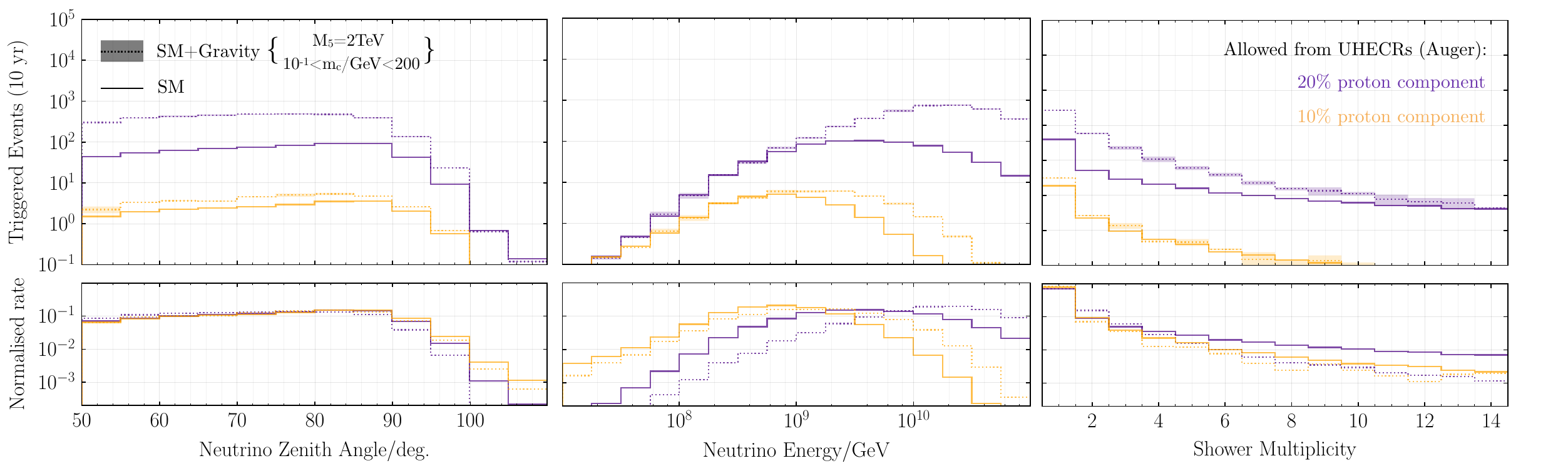}
\caption{Expected neutrino zenith (right), energy (center), and shower multiplicity (right) distributions after ten years of data taking with the IceCube-Gen2 radio array assuming two different GZK fluxes indicated with orange and purple colors. 
Continuous lines represent the SM prediction and shaded bands include Gravity-mediated interactions assuming $M_{5}=\SI{2}\TeV$ TeV and $m_{c}$ ranging from $\SI{0.1}\GeV$ to $M_{5}/10$.}
\label{fig:neutrino}
\end{figure*}

Figure~\ref{fig:neutrino} shows the expected number of triggered events as function of the neutrino zenith angle and energy when it enters in the detector. We also show the number of triggered showers per event. An enhancement of down-going events at high energies is observed. However, the energy deposition of this neutrinos is significantly smaller, resulting in an increase of low energy showers as shown in the main text. 

\section{Multiplicities per flavor} \label{appendix:norm_multiplicity}
\begin{figure*}
\centering
\includegraphics[width=1\textwidth]{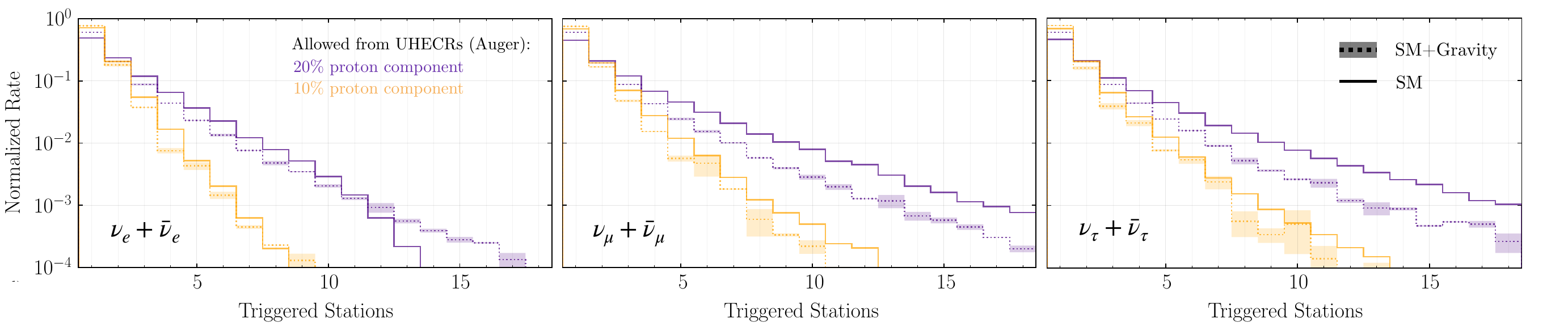}
\caption{Normalized distribution of the number of triggered stations after ten years of data taking with the IceCube-Gen2 radio array for $\nu_{e}$ (left), $\nu_{\mu}$ (center), and $\nu_{\tau}$ (right).
Two different GZK fluxes are indicated with orange and purple colors. Continuous lines represent the SM prediction and shaded bands include gravity-mediated interactions assuming $M_{5}=\SI{2}\TeV$ TeV and $m_{c}$ ranging from $\SI{0.1}\GeV$ to $M_{5}/10$.}
\label{fig:norm_multiplicity}
\end{figure*}

Radio detectors can not distinguish between different neutrino flavors. However, we think it is illustrative to present the number of trigger station for each neutrino type as shown in Figure~\ref{fig:norm_multiplicity}. It can be observed that in the case of $\nu_{e}$ interactions, the number of trigger stations is smaller than for $\nu_{\mu}$ or $\nu_{\tau}$.

\bibliographystyle{apsrev4-1}
\bibliography{main_text}

\end{document}